\documentclass[prd,10pt,twocolumn,superscriptaddress,floatfix,nofootinbib,preprintnumbers]{revtex4-1}
\usepackage[colorlinks=true,pdfstartview=FitV,breaklinks=true]{hyperref}

\usepackage{bm}
\usepackage[normalem]{ulem}
\usepackage{graphicx}
\usepackage{multirow}
\usepackage{soul}
\usepackage{amsmath}
\usepackage{fontawesome}
\usepackage{mathrsfs}
\usepackage{amssymb}
\usepackage{yfonts}
\usepackage{cancel}
\usepackage{color}
\usepackage{xspace}
\usepackage{slashed}
\usepackage{url}
\usepackage{verbatim}
\usepackage{mathtools}
\usepackage{upgreek}
\usepackage{units}
\usepackage{siunitx}
\usepackage{amstext}
\usepackage[sc]{mathpazo}
\usepackage{booktabs}
\usepackage{tabulary}
\usepackage{tabularx}
\usepackage{etoolbox}
\usepackage[version=4]{mhchem}
\usepackage[utf8]{inputenc}

\newcommand{\bq}{{\bf q}}
\newcommand{\bp}{{\bf p}}
\newcommand{\keV}{\text{ keV}}

\newcommand{\erg}{\rm erg}
\newcommand{\s}{\rm s}
\newcommand{\g}{\rm g}
\newcommand{\cm}{\rm cm}

\begin{document}

\title{Revisiting axion-electron bremsstrahlung emission rates in astrophysical environments}

\author{Pierluca Carenza}\email{pierluca.carenza@ba.infn.it}
\affiliation{Dipartimento Interateneo di Fisica “Michelangelo Merlin,” Via Amendola 173, 70126 Bari, Italy}
\affiliation{Istituto Nazionale di Fisica Nucleare - Sezione di Bari, Via Orabona 4, 70126 Bari, Italy}%

\author{Giuseppe Lucente}\email{giuseppe.lucente@ba.infn.it}
\affiliation{Dipartimento Interateneo di Fisica “Michelangelo Merlin,” Via Amendola 173, 70126 Bari, Italy}
\affiliation{Istituto Nazionale di Fisica Nucleare - Sezione di Bari, Via Orabona 4, 70126 Bari, Italy}%

\date{\today}
\smallskip
\begin{abstract}
The axion-electron coupling $g_{ae}$ is a generic feature of non-hadronic axion models. This coupling may induce
a variety of observable signatures, particularly in astrophysical environments. Here, we revisit the calculation of the axion-electron bremsstrahlung and provide a general formulation valid for a non-relativistic plasma with any level of degeneracy and for any axion mass. We apply our result to the Sun, red giant stars and white dwarfs. In particular, we prove that the approximations used to evaluate the axion emissivity in red giants agree with the exact result within $10\%$, comparable with other uncertainties in these studies.
In addition, this prescription allows the red giant and white dwarf bounds to be extended to massive axions.
 \end{abstract}
\maketitle

\section{Introduction}

\begin{figure}[t]
\centering
\includegraphics[width=0.4\textwidth]{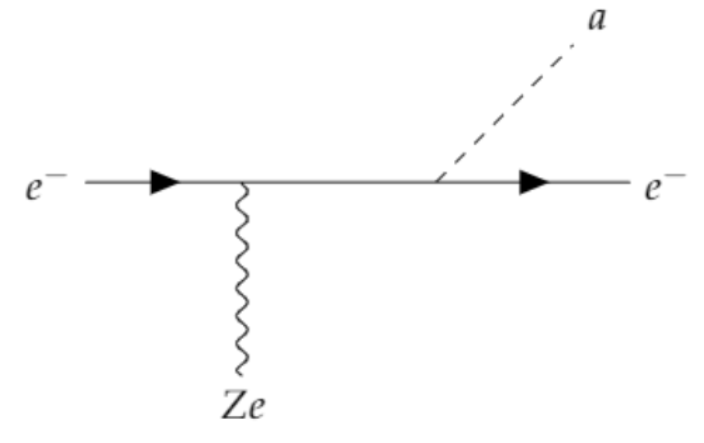}
\caption{Feynman diagram of the electron-ion bremsstrahlung. Note that a second amplitude with the vertices interchanged is not shown.}
\label{fig:brem}
\end{figure}

The axion-electron coupling naturally arises in non-hadronic axion models, like the Dine-Fischler-Srednicki-Zhitnitsky (DFSZ) \cite{Zhitnitsky:1980tq,Dine:1981rt}, and in more general theories, such as grand unified theories and string theory, predicting axion-like particles \cite{Cicoli:2012sz,Kachru:2003aw,Conlon:2006ur,Choi:2006qj,Arvanitaki:2009fg}. This coupling would give observable signals in both astrophysical context \cite{Raffelt:1985nj,Raffelt:1994ry} and laboratory experiments \cite{Riordan:1987aw,Bross:1989mp,Bjorken:1988as,Konaka:1986cb,Alves:2017avw,Bassompierre:1995kz,Scherdin:1991xy,Tsai:1989vw,Armengaud:2018cuy} (see Refs.~\cite{DiLuzio:2020wdo,Agrawal:2021dbo} for a recent review). 
In particular, light axions, with masses lower than the stellar temperature, can be efficiently produced in stars by electron bremsstrahlung (on ions or electrons) $e^{-}+Ze\rightarrow e^{-}+Ze+a$ (Fig.~\ref{fig:brem}), Compton scattering $e^{-}+\gamma\rightarrow e^{-}+a$, and electron-positron annihilation $e^{+}+e^{-}\rightarrow \gamma+a$. The last two processes  are important only in non-degenerate stars \cite{Pantziris:1986dc,Raffelt:1996wa}.
The extra energy-loss channel associated with axion emissivity would modify the stellar observables, giving the possibility to constrain their properties \cite{Isern:2008nt,Isern:2008fs,Viaux:2013lha,Bertolami:2014wua,Capozzi:2020cbu,Straniero:2020iyi} or explain possible hints of extra cooling in different stellar systems \cite{Giannotti:2015kwo,Giannotti:2017hny}. 
Stars in which electrons are more degenerate, such as the core of red giants (RGs) and white dwarfs (WDs) provide the most stringent bounds on the axion coupling with electrons.
Indeed, the RG bound excludes $g_{ae}\gtrsim 1.6\times10^{-13}$  \cite{Capozzi:2020cbu,Straniero:2020iyi} and the WD bound constrains $g_{ae}\gtrsim 2.8\times10^{-13}$ \cite{Isern:2008fs,Isern:2008nt,Bertolami:2014wua}. In these environments, the leading axion production channel is the electron-ion bremsstrahlung, while the electron-electron bremsstrahlung is suppressed by the electron degeneracy \cite{Raffelt:1996wa}. Given the relevance of bremsstrahlung processes in determining the axion emissivity in different stellar environments, we find it useful to revisit the previous calculation of this process including effects so far neglected in the literature, specifically
\begin{itemize}
\item[(i)] The effects of degeneracy. 
In the literature, the electron plasma is always assumed to be either completely degenerate or completely non-degenerate,  neglecting the cases of intermediate degeneracy. In our work we extend the calculation of the bremsstrahlung presenting the rate for any degree of degeneracy, relevant for a plethora of astrophysical environments.
We will show that this improved calculation agrees with the previous literature \cite{Raffelt:1996wa} within a few percent in the case of completely (non)-degenerate stars as the Sun and WDs.\\
Conversely, for RGs, our calculated axion emissivity results to be $\sim 25\%$ lower than the completely degenerate limit. 
This reduction is due to the partial degeneracy of the electron gas in the RG core, which was accounted in Refs.~\cite{Capozzi:2020cbu,Straniero:2020iyi} with an interpolation formula first proposed in Ref.~\cite{Raffelt:1994ry}. The interpolation formula is not needed in WDs since the plasma in a WD is typically more degenerate than in a RG star.

\item[(ii)] The effect of the axion mass. So far, all previous studies have assumed massless axions in the bremsstrahlung process. However, this assumption ceases to be valid if the axion mass $m_a$ becomes comparable to the stellar temperature $T$. In this case, the axion production would be Boltzmann suppressed and, consequently, the astrophysical bounds would be relaxed (see e.g. \cite{Carenza:2020zil,Lucente:2020whw}). 
Here, we quantify this behavior through a calculation which explicitly takes into account the axion mass in the matrix element of the electron-ion bremsstrahlung.
\end{itemize}

The plan of our work is as follows.
In Sec.~\ref{sec:production}, we discuss our revision of the axion production mechanism via bremsstrahlung. 
In Sec.~\ref{sec:applications}, we apply these results to representative astrophysical environments, namely, to the Sun, RGs, and WDs. In Sec.~\ref{sec:consequences}, we discuss in more detail the impact on the RG axion bound. Finally,  in Sec.~\ref{sec:conclusions} we conclude. Two Appendices follow. In Appendix~\ref{sec:Appendix}, we show the complete matrix element of the electron-ion bremsstrahlung and in Appendix~\ref{sec:appendixB} we summarize the results of the existing literature.

\section{Electron-ion bremsstrahlung}
\label{sec:production}

\begin{table*}[!t]
 \caption{
The emissivity per unit mass $\varepsilon_a$ evaluated in solar, RG, and WD conditions \cite{Raffelt:1996wa,Raffelt:2006cw}. For each environment, we show the typical density $\rho$, temperature $T$, the electron fraction $Y_e$, the electron degeneracy parameter $\eta$, the emissivity $\varepsilon_{\text{appr}}$ obtained through the approximate expressions in Ref.~\cite{Raffelt:1996wa} (see Appendix~\ref{sec:appendixB}), the emissivity $\varepsilon_{\text{ex}}$ through Eq.~\eqref{eq:emissivity}, and the discrepancy between them $(\varepsilon_{\text{appr}}-\varepsilon_{\text{ex}})/\varepsilon_{\text{appr}}$. In all the cases, we take as coupling constant $g_{ae}=10^{-13}$.}
\begin{center}
\begin{tabular}{lcccccccc}
\hline
Condition &$\rho$&$T$&$Y_{e}$&$\eta$&
$\varepsilon_{\text{appr}}$&$\varepsilon_{\text{ex}}$&$(\varepsilon_{\text{appr}}-\varepsilon_{\text{ex}})/\varepsilon_{\text{appr}}$\\
  &$(\g~\cm^{-3})$&$(\keV)$& &&
(erg~g$^{-1}$~s$^{-1}$)&(erg~g$^{-1}$~s$^{-1}$)& ($\%$) \\
\hline
\hline
Sun &$1.6\times10^{2}$&$1.3$&0.5&-1.72&$1.60\times 10^{-5}$&$1.58\times 10^{-5}$&1.1 \\
RG &$2\times 10^{5}$&$8.6$&0.5&6.16&$1.08$&$0.82$&24 \footnote{Using the interpolation formula in Eq.~\eqref{eq:interp}, this discrepancy is reduced to $\sim - 10\%$.}\\
WD &$2\times 10^{6}$&$1$&0.5&$2.14\times 10^{2}$ &$5.55\times 10^{-5}$&$5.55\times 10^{-5}$&0.02\\
\hline
\end{tabular}
\label{tab:eps}
\end{center}
\end{table*}

The axion interaction with electrons is described by the following Lagrangian \cite{Raffelt:1996wa}
\begin{equation}
    \mathcal{L}_{ae}=\frac{g_{ae}}{2 m_e} \bar{\psi}_{e}\gamma^{\mu}\gamma^{5}\psi_{e} \partial_\mu a\,,
\label{eq:lagrangian}
\end{equation}
where $\psi_{e}$ and $a$ are, respectively, the electron and axion fields, $m_e$ is the electron mass, and $g_{ae}$ is the dimensionless axion-electron coupling. \\
In the electron-ion bremsstrahlung, an electron is deflected by the electric field of a static ion and the final electron emits an axion. In this context, the Lagrangian in Eq.~\eqref{eq:lagrangian} is equivalent to
\begin{equation}
  \mathcal{L}_{ae}=  -i g_{ae}\bar{\psi}_{e}\gamma^{5}\psi_{e}\,a\,.
\end{equation}
We stress that this equivalence is not general; e.g. it ceases to be valid when two Goldstone bosons are attached to one fermion line.\\
The electron-ion bremsstrahlung matrix element is
\begin{equation}
\begin{split}
    &\mathcal{M}_{j}=\frac{g_{ae}\,Z_{j}e^{2}}{|\bq|(|\bq|^{2}+k_{S}^{2})^{1/2}}\\
    &\times\bar{u}(p_{f})\left[\gamma^{5}\frac{1}{\slashed{P}-m_{e}}\gamma^{0}+\gamma^{0}\frac{1}{\slashed{Q}-m_{e}}\gamma^{5}\right]u(p_{i})\,,
\end{split}
\label{eq:matel}
\end{equation}
where $u(p_{i}), u(p_{f})$ are the electron spinors, $p_{i}$, $p_{f}$ and $p_{a}$ are four-momentum of initial, final electrons and axion, $P=p_{f}+p_{a}$, $Q=p_{i}-p_{a}$, and $\bq=\bp_{f}+\bp_{a}-\bp_{i}$ is the momentum transfer. The term $[|\bq|(|\bq|^{2}+k_{S}^{2})^{1/2}]^{-1}$ is the Coulomb propagator in a plasma and $k_{S}$ is the Debye screening scale given by \cite{Raffelt:1985nk}
\begin{equation}
k_S^2 = \frac{4\pi \alpha \sum_j Z_j^2 n_j}{T}\,,
\end{equation}
where $n_j$ the number density of ions with charge $Z_j\,e$ and $\alpha$ is the fine structure constant.
The axion flux is found to be 
\begin{equation}
\begin{split}
    \frac{d^{2}n_{a}}{dt\,d\omega_{a}}=&2\pi\int\frac{2d^{3}\bp_{i}}{(2\pi)^{3}2E_{i}}\frac{2d^{3}\bp_{f}}{(2\pi)^{3}2E_{f}}\frac{|\bp_{a}|}{(2\pi)^{3}}\\
    &(2\pi)\delta(E_{i}-E_{f}-\omega_{a})\,|\mathcal{M}|^{2}f_{i}(1-f_{f})=\\
    =&\frac{1}{64\pi^{6}}\int d\cos\theta_{ia}\,d\cos\theta_{if}\,d\delta\,dE_{f}\\
    &|\bp_{i}||\bp_{f}||\bp_{a}||\mathcal{M}|^{2} f_{i}(1-f_{f})\,,
\end{split}
\label{eq:flux}
\end{equation}
where $\omega_{a}$, $E_{i}$ and $E_{f}$ are the energies of the axion, initial and final electrons respectively; $f_{i,f}$ are the electron distribution functions; $\theta_{ia}$, $\theta_{if}\in [0,\pi]$ are the angles between the initial electron and the axion and the final electron moments respectively; $\delta\in[0,2\pi]$ is the angle between the two planes determined by the vectors $\bp_i-\bp_a$ and $\bp_i-\bp_f$ and $|\mathcal{M}|^{2}=\frac{1}{4}\sum_{j} n_j \sum_{s} |\mathcal{M}_j|^{2}$ is the matrix element in Eq.~\eqref{eq:matel} averaged over the electron spins and summed over all the target ions. 
The calculation of this matrix element is non-trivial and we performed it with the help of the FeynCalc package \cite{Kublbeck:1990xc,Shtabovenko:2016sxi,Shtabovenko:2020gxv}. 
The complete result is shown in Appendix~\ref{sec:Appendix}. In the limit of vanishing axion mass, it reduces to 
\begin{equation}
\begin{split}
 &|\mathcal{M}|^{2}=\frac{1}{4}\sum_{j}n_{j}\sum_{\rm s}|\mathcal{M}_{j}|^{2}=\frac{g_{ae}^{2}e^{2}}{2}\frac{k_{S}^{2}T}{|\bq|^{2}(|\bq|^{2}+k_{S}^{2})}\\
 &\left[2\omega_{a}^{2}\frac{p_{i}\cdot p_{f}-m_{e}^{2}-K\cdot p_{a}}{(p_{i}\cdot p_{a})(p_{f}\cdot p_{a})}+2-\frac{p_{f}\cdot p_{a}}{p_{i}\cdot p_{a}}-\frac{p_{i}\cdot p_{a}}{p_{f}\cdot p_{a}}\right]\,,
  \end{split}
  \label{eq:matel2}
\end{equation}
where $K=p_{f}-p_{i}$. This result agrees with the literature \cite{Raffelt:1989zt} except for a sign in the term $\sim K\cdot p_{a}$. However, this difference is irrelevant for the results shown in Ref.~\cite{Raffelt:1989zt}, where $K\cdot p_{a}\simeq 0$.

\section{Applications to astrophysical environments}
\label{sec:applications}

We present here some applications of our results to  different astrophysical environments. Specifically, we will consider the Sun, RGs, and WDs. For simplicity, we characterize these environments assuming constant representative values for temperature $T$, density $\rho$, and electron fraction $Y_{e}$ \cite{Raffelt:1996wa,Raffelt:2006cw}. Typical values for these quantities in the environments we consider are given in Table~\ref{tab:eps}, together with the electron degeneracy parameter
\begin{equation}
    \eta=\frac{\mu -m_{e}}{T}\;,
\end{equation}
where $\mu$ is the electron chemical potential. An electron plasma is said to be non-degenerate if $\eta<0$ and degenerate if $\eta$ is larger than a few \cite{Raffelt:1996wa}.
For each source we calculate the emissivity, i.e. the energy emitted per unit mass and time, as
\begin{equation}
    \varepsilon_a=\frac{1}{\rho}\int_{m_a}^\infty d\omega_a\,\omega_a\,\frac{d^2n_a}{dtd\omega_a}\,,
\label{eq:emissivity}
\end{equation}
with ${d^2n_a}/{dt\,d\omega_a}$ from Eq.~\eqref{eq:flux}, taking by simplicity $Z=1$ and $n=\rho Y_e/m_N$, with $m_N=938$~MeV. 
In Table~\ref{tab:eps} we compare the massless axion emissivity ($\varepsilon_{\rm ex}$) with the literature in the suitable limit \cite{Raffelt:1996wa} ($\varepsilon_{\rm appr}$).
Note that we considered $Z=1$ for all the astrophysical environments. This assumption is valid as far as we are not interested in the absolute magnitude of the emissivity, but only in comparing two different formulas with the same input, as shown in Table~\ref{tab:eps}.
As discussed in Appendix~\ref{sec:appendixB}, in non-degenerate environments, as the Sun, we used the approximate formula in Eq.~\eqref{eq:ND}. Conversely, for strongly or mildly degenerate environments, we compared our result with Eqs.~\eqref{eq:D} and \eqref{eq:F}.

In the Sun, where electrons are supposed to be non-relativistic ($T\ll m_e$) and non-degenerate ($\eta<0$), the axion emissivity evaluated through Eq.~\eqref{eq:emissivity} ($\varepsilon_\text{ex}$) agrees with the existent non-degenerate limit in Eq.~\eqref{eq:ND} ($\varepsilon_{\rm appr}$) within $1\%$. We stress that even in this case the partial degeneracy plays a non-negligible role. In particular, if we ignore the Pauli blocking factor in Eq.~\eqref{eq:flux}, the emissivity becomes $\sim 6\%$ larger than the exact calculation, in agreement with Ref.~\cite{Hoof:2021mld}. However, the approximate expression in Eq.~\eqref{eq:ND} is given at the first order in $\kappa_S^2$, underestimating the full calculation in the completely non-degenerate limit. The combination of these two effects gives a $1\%$ discrepancy with respect to the exact calculation.
Thus, our result is useful for an accurate characterization of the solar axion flux, which needs a high level of precision \cite{Redondo:2013wwa,Jaeckel:2018mbn,Dafni:2018tvj,Hoof:2021mld}. Furthermore, it permits to extend these studies to massive axions. Note that a high-precision evaluation of the solar axion flux should not neglect the contribution of the electron-electron bremsstrahlung \cite{Redondo:2013wwa}. 

\begin{figure}
\centering
\includegraphics[width=0.45\textwidth]{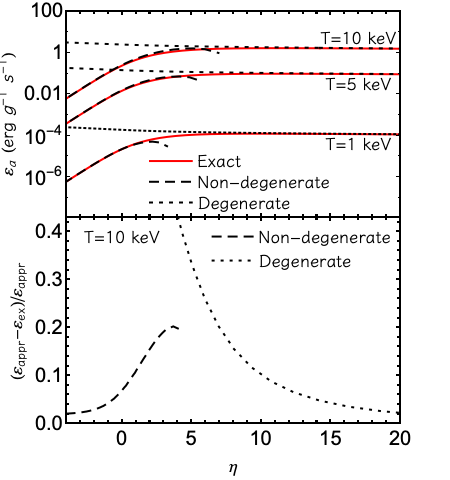}
\caption{Upper panel: comparison between the emissivity in Eq.~\eqref{eq:emissivity} (red solid line), the non-degenerate [Eq.~\eqref{eq:ND}, black dashed line] and the degenerate [Eq.~\eqref{eq:D}, black dotted line] approximation as a function of the electron degeneracy, at different values of the temperature $T$. The red curve interpolates between the dashed line and the dotted one for $0\lesssim\eta\lesssim10$.  Lower panel: discrepancy between the emissivity in Eq.~\eqref{eq:emissivity} ($\varepsilon_{ex}$) and the non-degenerate (black dashed line) and degenerate approximation (black dotted line) as a function of the degeneracy parameter $\eta$ at $T=10$~keV.}
\label{fig:deg}
\end{figure}

\begin{figure}
\centering
\includegraphics[width=0.45\textwidth]{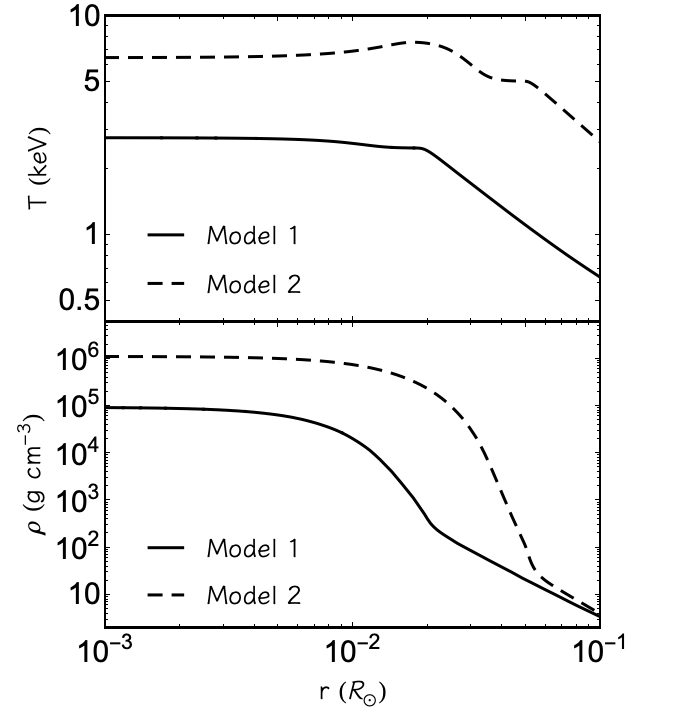}
\caption{Profiles of the RG temperature and density for Model 1 (solid line) and Model 2 (dashed line).}
\label{fig:profRG}
\end{figure}

\begin{figure*}[t!]
\centering
\includegraphics[width=0.45\textwidth]{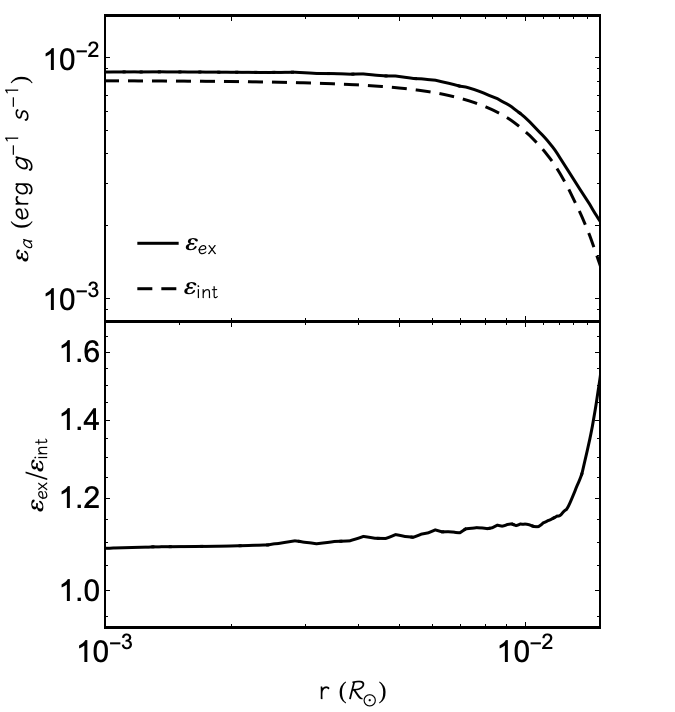}
\includegraphics[width=0.45\textwidth]{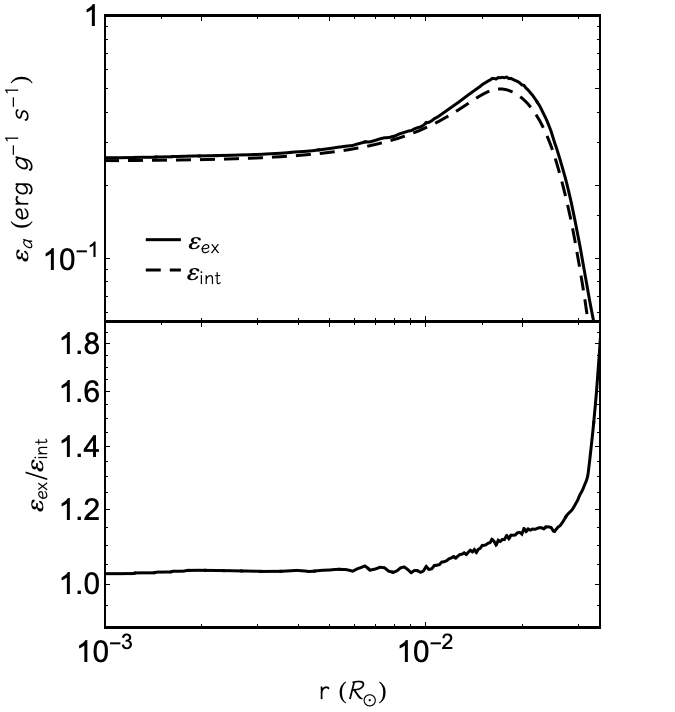}
\caption{Upper panels: the exact emissivity $\varepsilon_{\rm ex}$ (solid line) and the interpolated one $\varepsilon_{\rm int}$ (dashed line) as functions of the radius $r$ for Model 1 (left panel) and Model 2 (right panel). Lower panels: the ratio $\varepsilon_{\rm ex}/\varepsilon_{\rm int}$ as a function of the radius $r$ for Model 1 (left panel) and Model 2 (right panel).}
\label{fig:epsilonfig}
\end{figure*}

As the density increases, electrons become more degenerate. 
In RG cores, where $\eta\sim 6$, we find a discrepancy of $24\%$ with respect to the completely degenerate approximation. By increasing the temperature or lowering the density, this discrepancy increases. This behavior suggests that the discrepancy is related to the intermediate level of electron degeneracy so that the complete degenerate limit is not suitable in this situation. In particular, in the upper panel of Fig.~\ref{fig:deg}, we show the emissivity in Eq.~\eqref{eq:emissivity} (red solid line), the non-degenerate limit in Eq.~\eqref{eq:ND} (black dashed line), and the degenerate one in Eq.~\eqref{eq:D} (black dotted line) as a function of the electron degeneracy parameter $\eta$, for different temperatures in the range $1\,\keV\lesssim T\lesssim10\,\keV$. It is apparent that the non-degenerate approximation holds for $\eta\lesssim 0$, while the degenerate one is suitable for $\eta\gtrsim 10$. As depicted in the lower panel of Fig.~\ref{fig:deg}, at $T=10$~keV, the discrepancy between the emissivity evaluated through Eq.~\eqref{eq:emissivity} and the degenerate approximation is $\gtrsim 10\%$ at $5\lesssim\eta\lesssim 10$, in agreement with the result found for the typical RG conditions in Table~\ref{tab:eps}. Therefore, for $0\lesssim \eta \lesssim 10$, the exact calculation is needed to interpolate between the two regimes. A good strategy is proposed in Ref.~\cite{Raffelt:1994ry}, where in the intermediate regime the emissivity is evaluated as
\begin{equation}
 \varepsilon_{\rm int}^{-1}=\varepsilon_{ND}^{-1}+\varepsilon_{D}^{-1}\,,
 \label{eq:interp}
\end{equation}
which underestimates the axion emissivity by less than $10\%$ with respect to the emissivity in Eq.~\eqref{eq:emissivity}, where $\varepsilon_{ND}$ is given by Eq.~\eqref{eq:ND} ignoring screening effects and $\varepsilon_{D}$ by Eq.~\eqref{eq:D}.  This method is used in Refs.~\cite{Capozzi:2020cbu,Straniero:2020iyi} in order to evaluate the RG bound on $g_{ae}$, giving a more conservative result. We mention that in partial degenerate environments the electron-electron bremsstrahlung could give a subdominant, but non-negligible contribution. 

In WDs, in which electrons are extremely degenerate ($\eta\sim 200$), our calculation agrees remarkably well with the approximated one, with a discrepancy less than 0.1$\%$. We mention that in these very degenerate environments bremsstrahlung would be affected by crystallization (see e.g. Refs.~\cite{Nakagawa:1987pga,Nakagawa:1988rhp,Metcalfe:2004tb} for further details).

The emissivity in Eq.~\eqref{eq:emissivity} can be conveniently written as 
\begin{equation}
\begin{split}
    &\epsilon_{a}=4.73\,\erg~\g^{-1}~\s^{-1} \left(\frac{g_{ae}}{10^{-13}}\right)^{2}\left(\frac{k_{S}}{\keV}\right)^{2}\\
    &\left(\frac{T}{\keV}\right)^{3}\left(\frac{\rho}{\g ~\cm^{-3}}\right)^{-1}\mathcal{I}\left(\eta,\frac{m_{a}}{T},\frac{m_{e}}{T},\frac{k_{S}}{T}\right)\,,
\end{split}
\label{eq:epsint}
\end{equation}
where $m_a$ and $m_e$ are the axion and electron mass in $\keV$ and the function $\mathcal{I}$, evaluated using the D01GDF function of the NAG library, is tabulated for typical RG conditions and available in a public repository. 
\footnote{https://github.com/pierlucacarenza/Axion-Electron-Ion-Bremsstrahlung}\\
\section{Consequences on the Red Giant bound}
\label{sec:consequences}
\begin{figure}
\centering
\includegraphics[width=0.45\textwidth]{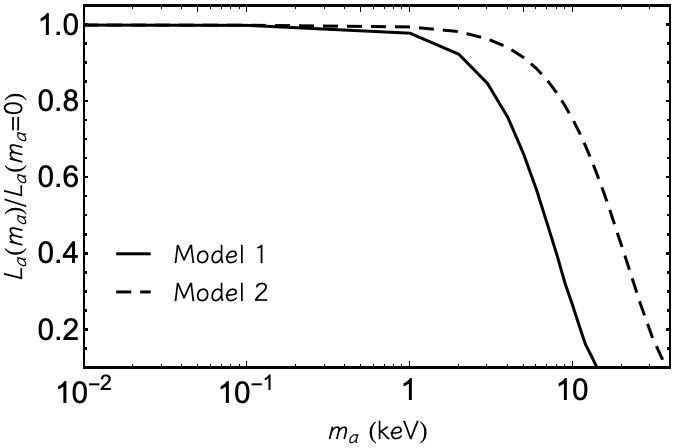}
\caption{The luminosity as a function of the axion mass for Model 1 (solid line) and Model 2 (dashed line).}
\label{fig:lamass}
\end{figure}

As discussed in the previous Section, the most important impact of the new computation of the electron-ion bremsstrahlung would be in RGs, affecting the high-precision axion bounds obtained in this environment. For this reason, in the following, we apply the formula in Eq.~\eqref{eq:epsint} to two RG models, computed by means of the Full Network Stellar evolution (FuNS) code~\cite{Straniero:2020iyi}. In these models, the core mass is $M=0.82~M_\odot$, the initial helium mass fraction is $Y=0.245$, and the metallicity is $Z=1.36\times10^{-3}$. The only difference between the two models is the age, $t_{\rm age}=10^{10.1073}$~yr for Model 1 and $t_{\rm age}=10^{10.1155}$~yr for Model 2, close to the RGB tip, the relevant evolutionary phase for axion bounds. In Fig.~\ref{fig:profRG}, we show the temperature $T$ (upper panel) and the density $\rho$ (lower panel) as functions of the radius $r$ in units of solar radius $R_\odot=6.96\times10^{5}$~km for our two models, which cover a large range of parameters for typical RG conditions. \\
In Fig.~\ref{fig:epsilonfig}, we compare the exact axion emissivity $\varepsilon_{\rm ex}$ in Eq.~\eqref{eq:emissivity} (solid line) with the interpolated one $\varepsilon_{\rm int}$ in Eq.~\eqref{eq:interp} (dashed line) for Model 1 (left panels) and Model 2 (right panels). In both the models, the interpolated emissivity underestimates the exact one by $\lesssim 10\%$ in the inner core $r\lesssim 10^{-2}$~$R_\odot$, while at larger radii the discrepancy increases since the density drops. As shown in the lower panels of Fig.~\ref{fig:epsilonfig}, in the innermost region of Model 1 the discrepancy is slightly larger than in Model 2 because the electron gas is less degenerate due to the lower density (see the lower panel in Fig.~\ref{fig:profRG}). In Model 2, the temperature peaks at $r\simeq 1.3\times10^{-2}~R_\odot$, reducing the electron degeneracy and causing an increase in both the emissivity and discrepancy.\\
Integrating the emissivity over the RG model, we obtain the luminosity $L_a$,
\begin{equation}
    L_a=4\pi\int dr\,r^2 \rho(r) \varepsilon_{a}\,.
\end{equation}
Since the emissivity is suppressed in the low-density region, in both the models the exact luminosity is larger than the interpolated one by $\sim10\%$, compatible with the discrepancy in the inner core. This is comparable to the theoretical and observational uncertainties discussed in Refs.~\cite{Capozzi:2020cbu,Straniero:2020iyi} (see e.g. Table 2 in \cite{Straniero:2020iyi}). Due to the larger emissivity, our revised calculation is expected to lead to a slightly stronger bound. However, its re-evaluation,  using our complete formula, is postponed to an upcoming paper~\cite{Oscar}.\\
In Fig.~\ref{fig:lamass} we show the luminosity as a function of the axion mass $m_a$ for the two models. As expected, the mass suppression begins at larger masses in Model 2, since the temperature is larger. In this way a RG bound for massive axions can be evaluated for the first time, using the complete matrix element in Eq.~\eqref{eq:matel} which takes into account the axion mass, always neglected in previous works. We postpone this analysis to a future work~\cite{Oscar}.

\section{Conclusions}
\label{sec:conclusions}

In this work we discussed the axion emissivity via electron-ion bremsstrahlung for any finite axion mass and any degree of electron degeneracy in a non-relativistic plasma, useful for different stellar environment conditions. This analysis follows the recent efforts in calculating precisely the astrophysical axion fluxes, which require a more accurate understanding of the relevant axion production mechanisms.
Our result gives an emissivity in the massless case which agrees with the literature \cite{Raffelt:1996wa} within $\sim 5\%$ in the Sun and WDs. The largest difference is found in RGs, where our result is $\sim 25\%$ lower than the completely degenerate limit and this difference is due to the intermediate electron degeneracy. 
In this regime, the interpolation formula in Ref.~\cite{Raffelt:1994ry}, used in the evaluation of the RG bound \cite{Capozzi:2020cbu,Straniero:2020iyi}, underestimates the axion emissivity by less than $10\%$ with respect to our complete calculation, leading to a more conservative bound. Since the uncertainties described in Refs.~\cite{Capozzi:2020cbu,Straniero:2020iyi} have a similar magnitude, the impact of our revised calculation can only be assessed with a detailed re-evaluation of the RG bound, to be discussed in a forthcoming work~\cite{Oscar}. Besides the consequences on the massless axion limit, our result allows the extension of the RG and WD bounds to axion masses larger than the stellar temperature~\cite{Oscar}. These bounds should be compared to the existing experimental bounds in the same axion mass region from EDELWEISS III \cite{Armengaud:2018cuy} and GERDA \cite{GERDA:2020emj}. Notice, however, that bounds based on stellar energy loss are completely independent on the assumption that axions constitute the totality of the dark matter in the universe.

\section*{Acknowledgement}
We warmly thank Maurizio Giannotti, Joerg Jaeckel, M.C. David Marsh,  Alessandro Mirizzi, and Georg Raffelt for comments on the paper and useful discussions. We acknowledge Oscar Straniero for providing the stellar data.
The work of P.C. and G.L. is partially supported by the Italian Istituto Nazionale di Fisica Nucleare (INFN) through
the “Theoretical Astroparticle Physics” project and by
the research grant number 2017W4HA7S “NAT-NET:
Neutrino and Astroparticle Theory Network” under the
program PRIN 2017 funded by the Italian Ministero dell’Università e della Ricerca (MUR).

\appendix
\section{Complete matrix element}
\label{sec:Appendix}

The complete matrix element of the electron-ion bremsstrahlung is
\begin{equation}
\begin{split}
  &|\mathcal{M}|^2=
  g_{ae}^{2}e^{2}\frac{k_{S}^{2}T}{|\bq|^{2}(|\bq|^{2}+k_{S}^{2})}\\
  &\frac{1}{(2(p_a\cdot p_f)+m_a^2)^2\,(m_a^2-2(p_a\cdot p_i))^2}\\
  &\big[4((p_a\cdot p_f)^2(4(p_a\cdot p_i)^2+m_a^2(p_f\cdot p_i-m_e^2+\\
  &+2\,E_i(\omega_a-E_f))-4(p_a\cdot p_i)(m_a^2+\omega_a(E_i-E_f)))+\\
  &+2(p_a\cdot p_f)(2(m_a^2+\omega_a(E_i-E_f))(p_a\cdot p_i)^2+\\
  &-(p_a\cdot p_i)^3-m_a^2(m_a^2-\omega_a^2)(m_e^2-p_f\cdot p_i)+\\
  &-(p_a\cdot p_i)((m_a^2-2\omega_a^2)(p_f\cdot p_i)+m_a^4-m_a^2\,m_e^2+\\
  &+m_a^2(3\omega_a(E_i-E_f)-2E_f\,E_i)+2m_e^2\omega_a^2)+\\
  &+m_a^4\,E_i(\omega_a-2E_f))+\\
  &+m_a^2(-(p_a\cdot p_i)^2(-p_f\cdot p_i+m_e^2+2E_f(\omega_a+E_i))+\\
  &-m_a^2(m_a^2-\omega_a^2)(m_e^2-p_f\cdot p_i)+\\
  &+2(p_a\cdot p_i)(((\omega_a^2-m_a^2)(p_f\cdot p_i)+m_e^2(m_a^2-\omega_a^2)+\\
  &+m_a^2\,E_f(\omega_a+2E_i))-2m_a^4E_f\,E_i)+\\
  &-2(p_a\cdot p_i)(p_a\cdot p_f)^3)\big]\,.
\end{split}
\end{equation}

\section{Summary of the literature}
\label{sec:appendixB}

\begin{figure}[t!]
\centering
\includegraphics[width=0.45\textwidth]{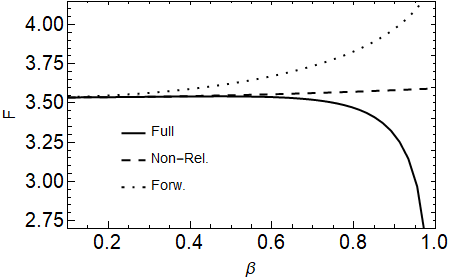}
\caption{Comparison between the function F in Eq.~\eqref{eq:F} (solid line) and the non-relativistic (Eq.~\eqref{eq:RNR}, dashed line) and forward (Eq.~\eqref{eq:RR}, dotted line) approximations as function of the electron velocity.}
\label{fig:Fapprox}
\end{figure}

In the following, we present and discuss the approximate formulas used in the literature in the completely (non)-degenerate limits. Electrons are assumed to be non-degenerate when the degeneracy parameter $\eta$ is smaller than a few and degenerate otherwise.\\
In the non-degenerate limit, the emissivity is expressed as an expansion to the first order in $k_{S}$ \cite{Raffelt:1996wa},
\begin{equation}
 \varepsilon_{ND}=g_{ae}^{2}\sqrt{\frac{2}{\pi}}\frac{32}{45} \frac{\alpha^{2}n_{e}T^{2.5}}{m_{e}^{3.5}\rho} \left(1-\frac{5k_{S}^{2}}{8m_{e}T}\right)\left(\sum_{j} n_j\,Z_j^2\right)\,,
 \label{eq:ND}
\end{equation}
where $n_{e}$ is the electron density and $\rho$ is the stellar density. In the degenerate limit, the emissivity can be written as \cite{Raffelt:1996wa}
\begin{equation}
\varepsilon_{D}=g_{ae}^2\frac{\pi \alpha^2}{60\, \rho\, m_e^2}\, \left(\sum_{j} n_j\,Z_j^2\right) F\,,
\label{eq:D}
\end{equation}
where the function $F$ is
\begin{equation}
\begin{split}
&F=\frac{2\pi}{16\pi^2}\int_{-1}^{1} dc_{if}  \int_{-1}^{1} dc_{ia} \int_0^{2\pi} d\delta \\ &\frac{(1-\beta^2)\left[2(1-c_{if})-(c_{ia}-c_{fa})^2\right]}{(1-c_{ia}\beta)(1-c_{fa}\beta)(1-c_{if})(1-c_{if}+\kappa^2)}\,,
\end{split}
\label{eq:F}
\end{equation}
with $c_{if}$ the cosine of the angle between $\textbf{p}_i$ and $\textbf{p}_f$, $c_{ia}$ the cosine of the angle between $\textbf{p}_i$ and $\textbf{p}_a$, $c_{fa}=c_{ia}\,c_{if}+\sqrt{1-c_{ia}^2}\sqrt{1-c_{if}^2}\cos\delta$, $\beta$ is the electron velocity, $p_{F}$ the electron Fermi momentum, and $\kappa^2=k_S^2/2\,p_F^2$. As discussed in Sec.~\ref{sec:applications}, the exact calculation in Eq.~\eqref{eq:emissivity} is needed to interpolate between these two regimes, for $0\lesssim \eta\lesssim 10$. However, a good approach in agreement within $10\%$ with the exact calculation is the interpolation formula suggested in Ref.~\cite{Raffelt:1994ry} $\varepsilon^{-1}=\varepsilon_{ND}^{-1}+\varepsilon_{D}^{-1}$. \\

In addition, in the case of a degenerate and non-relativistic electron plasma, Eq.~\eqref{eq:F} is simplified by expanding the integrand to the first order in the $\beta$, and after an integration one obtains
\begin{equation}
    F=\frac{2}{3}\ln\left(\frac{2+\kappa^2}{\kappa^2}\right)+\left[\frac{2+5\kappa^2}{15}\ln\left(\frac{2+\kappa^2}{\kappa^2}\right)-\frac{2}{3}\right]\,\beta^{2}\,.
    \label{eq:RNR}
\end{equation}
As shown in Fig.~\ref{fig:Fapprox}, the non-relativistic approximation reported in Eq.~\eqref{eq:RNR} (dashed line) accurately reproduces the exact calculation in Eq.~\eqref{eq:F} (solid line) up to $\beta\lesssim 0.6$.
In the literature, a further approximation is employed for a relativistic plasma.
Observing that the Coulomb scattering is mostly forward, i.e. the main contribution to the integral Eq.~\eqref{eq:F} is from $c_{if}\simeq 1$, $c_{fa}=c_{ia}$ is assumed only in the denominator of Eq.~\eqref{eq:F}, obtaining
\vspace{1cm}
\begin{equation}
\begin{split}
&F=\frac{2}{3}\ln\left(\frac{2+\kappa^2}{\kappa^2}\right)+\left[\frac{2+3\kappa^2}{6}\ln\left(\frac{2+\kappa^2}{\kappa^2}\right)-1\right]\,f(\beta)\,,\\
&f(\beta)=\frac{3-2\beta^2}{\beta^2}-\frac{3(1-\beta^2)}{2\beta^3}\ln\left(\frac{1+\beta}{1-\beta}\right)\,.
\end{split}
  \label{eq:RR}
\end{equation}
As can be seen from Fig.~\ref{fig:Fapprox}, the forward approximation in Eq.~\eqref{eq:RR} (dotted line) is not a good approximation at all, especially in the relativistic limit, where it is usually used. Indeed, as $\beta$ approaches to one, Eq.~\eqref{eq:F} goes to zero because of the $1-\beta^{2}$ term in the numerator. 
To summarize, Eq.~\eqref{eq:F} cannot be approximated in the relativistic limit as Eq.~\eqref{eq:RR} and the integral must be numerically evaluated. 

\bibliographystyle{bibi.bst}
\bibliography{biblio.bib}

\end{document}